\documentclass[10pt,twocolumn,english,twocolumn]{revtex4-1}
\usepackage{amssymb}
\usepackage{graphicx}

\makeatletter

\usepackage{lmodern}
\usepackage[T1]{fontenc}

\makeatother

\usepackage{babel}
\begin{document}

\title{All Optical Stabilization of a Soliton Frequency Comb in a Crystalline
Microresonator }

\author{J. D. Jost\textsuperscript{1,{$\dagger$}},E. Lucas\textsuperscript{1,{$\dagger$}}, T. Herr\textsuperscript{1,2},
C. Lecaplain\textsuperscript{1}, V. Brasch\textsuperscript{1}, M.
H. P. Pfeiffer\textsuperscript{1}, T. J. Kippenberg\textsuperscript{1}}
\email{tobias.kippenberg@epfl.ch}

\affiliation{1. \'{E}cole Polytechnique F\'{e}d\'{e}rale de Lausanne (EPFL), CH-1015 Lausanne,
Switzerland.\\2. Centre Suisse d'Electronique et de Microtechnique
(CSEM), Ne\^{u}chatel, Switzerland }

\begin{abstract}
Microresonator based optical frequency combs (MFC) have demonstrated promise in extending the capabilities of optical frequency combs. Here we demonstrate all optical stabilization of a low noise temporal soliton based MFC in a crystalline resonator via a new technique to control the repetition rate. This is accomplished by thermally heating the microresonator with an additional probe laser coupled to an auxiliary optical resonator mode. The offset frequency is controlled by stabilization of the pump laser frequency to a reference optical frequency comb. We analyze the stabilization by performing an out of loop comparison and measure the overlapping Allan deviation. This all optical stabilization technique can prove useful as a low added noise actuator for self-referenced microresonator frequency combs. 
\end{abstract}

\maketitle

Since their inception \cite{Jones2000}, optical frequency combs (OFC)
have significantly changed the way precise optical frequency measurements
are performed. Typically created from a train of ultra short pulses
emitted by mode-locked lasers in combination with supercontinuum generation
an equidistant grid of optical frequency markers is generated. Any
of the frequency components can be written in terms of the expression
$f_{\mathrm{n}}=n\cdotp f_{\mathrm{rep}}+f_{\mathrm{ceo}}$. Where
$n$ is an integer and parameters $f_{\mathrm{rep}}$ and $f_{\mathrm{ceo}}$
correspond to the repetition rate of the pulse train and to the carrier
envelope offset frequency. Discovered in 2007, microresonator based
optical frequency combs (MFCs) \cite{DelHaye2007,Kippenberg2011}
generated via parametric frequency conversion \cite{Kippenberg2004a,Savchenkov2004}
opened up a new parameter space for OFCs, achieving compact form factor
and on chip integration, enabling high repetition rates (typically
$>10$ GHz), and broad spectral from the near to the mid-infrared.
Such large mode spacing in the microwave regime \cite{Kippenberg2011}
is beneficial in applications such as astronomical spectrometer calibration
\cite{Steinmetz2008}, dual comb coherent Raman imaging \cite{Ideguchi2013},
high speed optical sampling and optical telecommunication \cite{Pfeifle2014}. 

A detailed understanding of how frequency comb formation arises in
these systems has been obtained in parallel by both experimental \cite{Herr2012}
and theoretical work \cite{Coen2013,Lamont2013,Chembo2010}. To generate
a MFC, a continuous wave laser is tuned into a cavity resonance and
once the intracavity intensity is above the parametric threshold frequency
comb formation begins via degenerate and non degenerate four-wave
mixing and many possible states can result. Multiple sub-combs are
formed when the variation in the cavity free spectral range is small
compared to the cavity linewidth. These states exhibit many types
of noise \cite{Herr2012}, but can be brought into the low noise regime
via $\delta-\Delta$ matching \cite{Herr2012}, parametric seeding
\cite{DelHaye2014}, injection locking \cite{Li2012,DelHaye2014},
or via the observation of mode-locking \cite{Saha2012a}. In the opposite
regime it is possible to form low phase noise states for narrow bandwidth
combs \cite{Herr2012,Ferdous2011,Papp2011}. An additional route that
is used in this work is temporal dissipative cavity soliton formation
\cite{Herr2013,Leo2010}. Temporal dissipative soliton formation results
in deterministic low noise, smooth spectral envelope frequency combs
that in time domain give rise to ultrashort optical pulses. In this
manner sub 200 fs pulses were generated in a crystalline microresonator
using a single soliton state, and in conjunction with external broadening
it was possible to determine $f_{\mathrm{ceo}}$ via a 2f-3f self-referencing
\cite{Jost15}. 

\begin{figure}[t]
\centering{}\includegraphics[width=0.9\columnwidth]{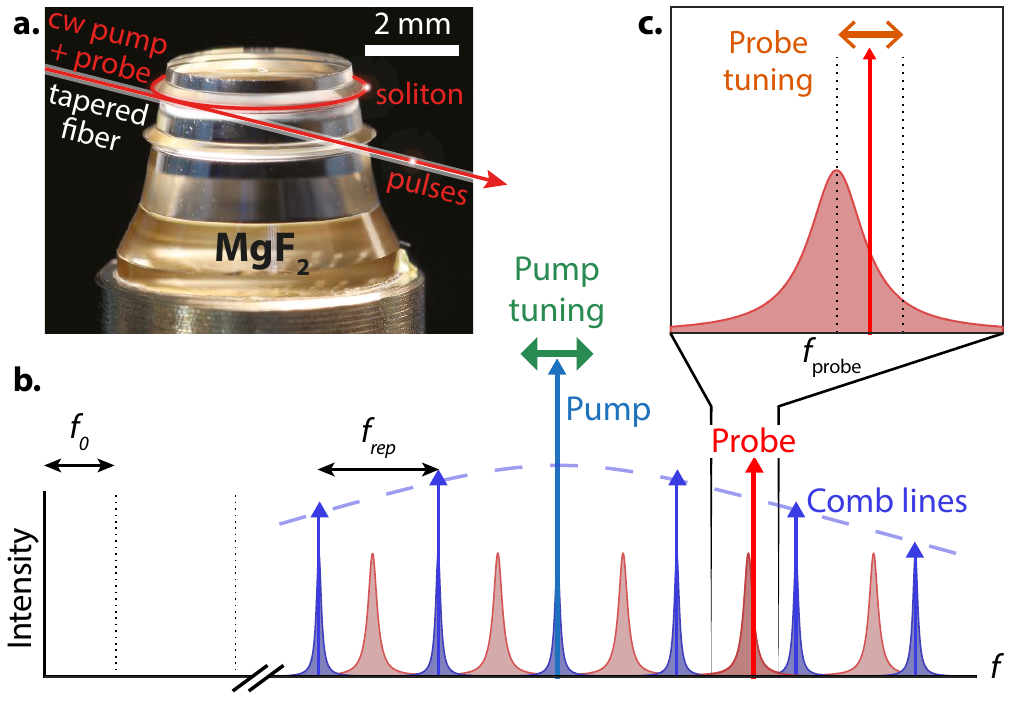}\protect\caption{Microresonator and actuation scheme: \textbf{a.} The $\mathrm{MgF_{2}}$
WGM microresonator is formed by the top protrusion going around the
circumference (the lower one is not used in this work). A tapered
optical fiber couples in and out the pump and probe lasers and a portion
of the generated soliton. \textbf{b.} Schematic of the basic concept
used to stabilize the repetition rate. The CW pump laser (blue) is
tuned into resonance of the high Q modes that produces the soliton
frequency comb producing mode (blue). Tuning the pump frequency changes
$f_{\mathrm{o}}$. The auxiliary lower-Q spatial mode family of the
cavity is shown in red. The probe laser is tuned to the high frequency
side of one of the resonances. \textbf{c.} A zoom in of the relevant
auxiliary mode. Tuning the probe laser in the resonance changes the
power coupled in and thus the FSR for both modes and changes $f_{\mathrm{rep}}$.
\label{fig:Stabilization-scheme}}
\end{figure}

The ability to self-reference an OFC is one of the key requirements
of a MFC being useful for precision metrology. In some applications
of MFCs such as atomic clocks \cite{Papp2014} it is important to
also stabilize both $f_{\mathrm{rep}}$ and $f_{\mathrm{ceo}}$. It
has been well established \cite{DelHaye2008} that it is possible
to independently control both parameters. One unique aspect of MFCs
is that the pump laser ($f_{\mathrm{o}}$) is the central comb line
and changing its frequency changes $f_{\mathrm{ceo}}$. This can also
affect the repetition rate via thermal and Kerr effects but an orthogonal
basis can be found with other control parameters \cite{DelHaye2008}.
Controlling the repetition rate is possible through a variety of means:
changing the pump power \cite{DelHaye2008}, actuating on a piezo
electric crystal that is in contact with the resonator \cite{Papp2012},
or heating or cooling the whole system \cite{Savchenkov2013a}. Here
we demonstrate all optical stabilization of both $f_{\mathrm{rep}}$
and $f_{\mathrm{ceo}}$ of a soliton based MFC by controlling $f_{\mathrm{o}}$
via the pump laser detuning and a new optical technique for the control
of $f_{\mathrm{rep}}$. For this, an additional probe laser is coupled
into a different cavity mode than the ones used for frequency comb
generation. By adjusting the detuning of the probe laser, the power
coupled into the microresonator is changed affecting $f_{\mathrm{rep}}$
(see figure \ref{fig:Stabilization-scheme}). Auxiliary modes have
also been used to monitor the resonator temperature and thus the repetition
rate \cite{Strekalov2011} and to compensate for thermal nonlinear
effects \cite{Grudinin2011}.

\begin{figure*}
\centering{}\includegraphics[width=0.9\textwidth]{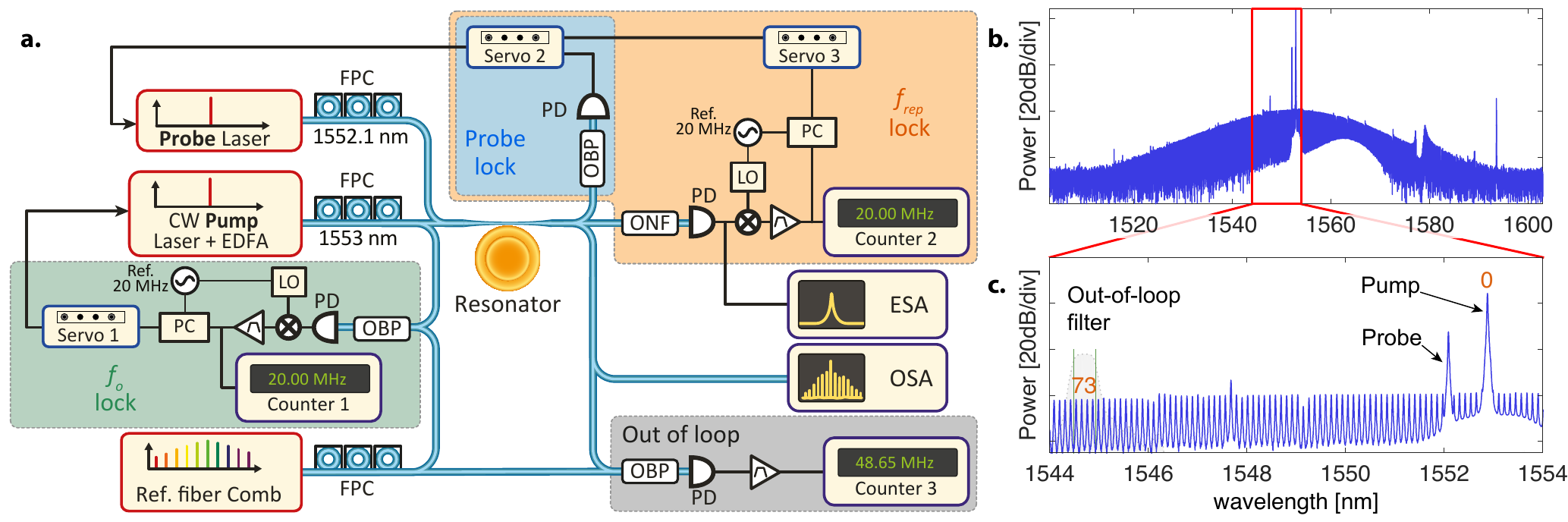}\protect\caption{All optical stabilization of a soliton frequency comb. \textbf{a.}
Experimental setup for generating and stabilizing the soliton frequency
comb. The details are provided in the text. The components are: continuous
wave (CW) pump and probe lasers, erbium doped fiber amplifier (EDFA),
fiber polarization controller (FPC), optical bandpass filter (OBP),
optical notch filter (ONF), optical spectrum analyzer (OSA), photodetector
(PD), local oscillator (LO), phase comparator (PC), electrical spectrum
analyzer (ESA)\label{fig:Experimental-Setup}\textbf{ b.} The generated
optical spectrum as measured on the OSA. \textbf{c.} A zoom in of
the optical spectrum. The pump laser and the probe laser's position
is shown. Due to the limited resolution of the OSA the probe appears
merged with the 7th comb line. The position of the out-of-loop optical
filter, covering several comb lines of both frequency combs, is shown
overlaid on the spectrum. It is centered \textasciitilde{} 73 comb
lines away from the pump laser (0th comb line). \label{fig:Single-Soliton-Spectrum}}
\end{figure*}

Whispering gallery mode microresonators (WGM) support azimuthally
symmetric optical modes, and in this work the modes are confined in
a small protrusion around an axially symmetric crystalline magnesium
fluoride ($\mathrm{{MgF_{2}}}$) microresonator \cite{Ilchenko2004,Herr2013}.
The geometry of the resonator, along with the material properties
determine the number and type of optical modes that can be supported.
It is possible to engineer the protrusion to support only one mode
\cite{Grudinin2015}. However in this work we leverage having more
than one spatial mode in the resonator for stabilization of the repetition
rate. One advantage of crystalline resonators is the exceptionally
high quality factors (Q) exceeding $10^{10}$ \cite{Braginsky1989,Grudinin2006}
that can be obtained. In this work the resonance used for OFC generation
has a $Q\sim10^{9}$ and a free spectral range of $\sim14.09$ GHz.
To form the OFC, continuous wave laser light from a 1553 nm fiber
laser (pump laser) with 240 mW is coupled evanescently into the resonator
via a tapered optical fiber \cite{Spillane2003}. Single or multiple
soliton states are generated in the resonator by performing a controlled
frequency scan in the direction of decreasing frequency with respect
to the resonance , and by choosing the correct stopping frequency
different soliton states can be accessed \cite{Herr2013}. A portion
of the light is coupled out via the tapered optical fiber, and the
resulting spectrum for a single soliton can be seen in figure~\ref{fig:Experimental-Setup}. 

The principle of the stabilization scheme for $f_{\mathrm{rep}}$
and $f_{\mathrm{o}}$ is presented in figure~\ref{fig:Stabilization-scheme}.
$f_{\mathrm{o}}$ is stabilized by offset locking the pump laser to
an external reference \cite{DelHaye2008}. Actuating on $f_{\mathrm{rep}}$
is done via a second ``probe'' laser at $\thicksim1552.1$~nm with
$\thicksim2$ mW, coupled and locked to a resonance of another mode
family with lower Q. By changing the lock point in this resonance,
more or less optical power is coupled into the resonance. Due to absorption
losses in the cavity this heats the resonator, changing the overall
size and refractive index and thus the repetition rate without loss
of the soliton state. In addition the Kerr effect also changes the
effective index of refraction on fast time scales. However, the magnitude
of this effect is negligible compared to the thermo-refractive index
change. 

The relevant stages of the experimental setup for the stabilization
of the soliton frequency comb are presented in figure~\ref{fig:Experimental-Setup}.
The locking setup for $f_{\mathrm{o}}$ is shown in the green panel.
A portion of the pump light is heterodyned with light from a reference
comb consisting of a commercial self-referenced and stabilized fiber
based optical frequency comb with a repetition rate of $\thicksim250$~MHz.
An optical bandpass filter (OBP) is used to filter (\textasciitilde{}
50 GHz bandwidth) the light preventing saturation of the photodetector
(PD). The heterodyning produces multiple beat signals between the
pump laser and the fiber frequency comb, which represent the frequency
offset of the pump with the near by comb lines. One of the beat signals
is electronically mixed to 20 MHz with a local oscillator phased locked
to the master oscillator (commercial atomic clock). A portion of the
signal is sent to a frequency counter. The rest is sent to a digital
phase comparator (PC) with a 20 MHz reference, which outputs an error
signal to a proportional and integral (PI) servo controller (servo~1).
This provides feedback to change frequency of the pump fiber laser
via an internal piezoelectric actuator. 

The stabilization setup for $f_{\mathrm{rep}}$ is shown in the orange
box in figure~\ref{fig:Experimental-Setup}a. The soliton pulse leaving
the resonator sits on a strong continuous wave background, which is
partially filtered out using an optical notch filter (ONF \textasciitilde{}
50 GHz bandwidth) to prevent saturation of the subsequent high speed
PD. To control $f_{\mathrm{rep}}$, the additional probe fiber laser
is coupled into the same tapered optical fiber as the pump laser,
and its frequency is tuned into a different spatial optical mode resonance
of the microresonator. A portion of the light coming out of the resonator
is sent to the probe locking setup depicted in the blue box in figure~\ref{fig:Experimental-Setup}a.
An OBP filter is used to pass primarily the transmitted probe light
to a photodetector. The probe laser's frequency is then locked to
the high frequency side of the cavity resonance, where thermal locking
is also supported. To suppress nonlinearities, lower powers and a
lower Q mode are used. It was not necessary to determine the exact
mode of the cavity, because many different modes showed control over
$f_{\mathrm{rep}}$. A PI servo controller with an adjustable set
point (servo~2) controls the lock point of the probe laser on the
side of the cavity resonance. The servo controller feeds back to a
piezoelectric actuator on the probe fiber laser, which adjusts the
laser's frequency. To fix $f_{\mathrm{rep}}$ an additional control
signal is needed which is derived from the repetition rate heterodyne
signal described above. This signal is down mixed from 14.094 GHz
to 20 MHz using a local oscillator that is referenced to the master
oscillator. The signal is filtered with an electronic bandpass filter
and a portion is sent to a counter. The rest is sent to a digital
PC with the 20 MHz signal from the master oscillator where any phase
error produces an error signal for a PI servo controller (servo~3).
The correction signal from the output of servo 3 is input to the servo
2, which maintains the probe laser's lock on the probe resonance,
adjusting the lock point detuning of the probe laser to maintain the
repetition rate phase lock. The time constants of the systems were
not directly measured but rather the appropriate proportional and
integral time constants were experimentally determined. It should
be noted that a sub 1 s integral time constant was used for the $f_{\mathrm{rep}}$
servo controller due to the observed slow response of the system,
indicating that thermal effects appear to dominate the response. To
verify the stabilization is not injecting significant noise an independent
out-of-loop measurement is performed by heterodyning a portion of
the generated soliton comb with the reference fiber based comb (gray
box in figure~\ref{fig:Experimental-Setup}). An OBP filter is used
to select a portion of both frequency combs centered at 1544.6 nm.
Due to the \textasciitilde{} 50 GHz bandwidth width of the filter
multiple heterodyne beat signals are observed. Using a tunable electronic
filter, one signal $f_{\mathrm{ol}}$ is selected and sent to a frequency
counter. 

\begin{figure}
\begin{centering}
\includegraphics[width=0.9\columnwidth]{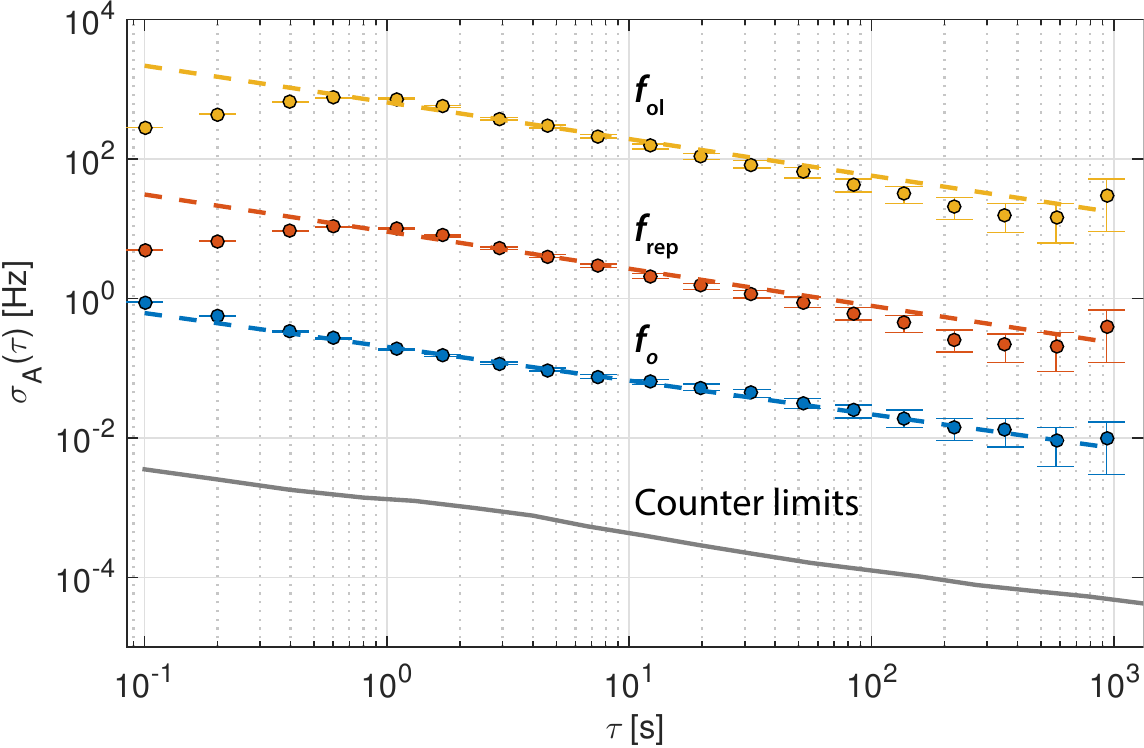}
\par\end{centering}

\protect\caption{Overlapping Allan deviation for a stabilized single soliton state.
Counting data with 100 ms gate time was taken for 3086 s. The resulting
OAD is shown by the traces for the pump frequency (blue), repetition
rate (red), and the independent out-of-loop measurement (yellow).
The error bars are one standard deviation. The dashed lines are a
linear fit for the data $\geq$1 s. The gray trace shows the result
of counting the 20 MHz master oscillator used in the experiment. \label{fig:Allan-Deviation-of-Single}}
\end{figure}

We analyze the stability of the full system by performing Allan deviation
measurements on $f_{\mathrm{rep}}$, $f_{\mathrm{o}}$, and $f_{\mathrm{ol}}$
on both a single and multiple soliton states. The heterodyne beat
frequencies are measured on Hewlett Packard 53131A high resolution
counters and the overlapping Allan deviation (OAD) is processed from
the recorded frequency series. For the single soliton state shown
in figure~\ref{fig:Single-Soliton-Spectrum}b counting data was taken
for 3086 s at 100 ms gate time, yielding the OAD plot in figure~\ref{fig:Allan-Deviation-of-Single}.
At 100 ms the absolute fluctuations of the 14.094~GHz rep rate beat
are $\sigma_{\mathrm{A,rep}}=4.98$~Hz corresponding to a fractional
deviation of $3.5\times10^{-10}$. The absolute frequency fluctuations
of $f_{\mathrm{o}}$ at 100 ms are $\sigma_{\mathrm{A,o}}=8.9\times10^{-1}$~Hz.
Only the absolute fluctuations of $f_{\mathrm{o}}$ can be measured
since the exact value of $f_{\mathrm{o}}$ is not known in this experiment.
For out-of-loop measurement the beat frequency was 70.3~MHz and the
absolute fluctuations at 100 ms were found to be $\sigma_{\mathrm{A,ol}}=283$~Hz.
The bump in $\sigma_{\mathrm{A,rep}}$ and $\sigma_{\mathrm{A,ol}}$
is probably a result of the slow thermal response of the system. Performing
a fit to the slopes, using the data for times $\geq1$s to make sure
the system was stabilized, we find that $f_{\mathrm{rep}}$, $f_{\mathrm{o}}$
and $f_{\mathrm{ol}}$ average down like $\tau_{\mathrm{rep}}^{-0.53}$,
$\tau_{\mathrm{o}}^{-0.48}$ and $\tau_{\mathrm{ol}}^{-0.53}$. We
observe here a dependence of $\sim\tau^{-0.5}$ while for a phase
locked system the expected dependence is rather $\sim\tau^{-1}$ \cite{Bernhardt2009a}.
This observation is related to the properties of the Hewlett Packard
53131A counters used. They are a $\Lambda$-type counter because they
perform a weighted average of the frequency over the gate time to
enhance the resolution. As a result, the computed deviation differs
from the true OAD but still measures the stability of the system \cite{Dawkins2007}.
Also, this counter cannot be read without dead time between consecutive
measurements. This creates a bias \cite{Lesage1983} in the processed
OAD for the phase-locked system and leads to a $\tau^{-0.5}$ dependence
in the presence of white phase noise. In our case the decreasing behavior
of the fluctuations for longer gate time still shows the stability
transfer of the master oscillator to the soliton frequency comb. The
out-of-loop measurement is approximately a factor $\sim71.5$ higher
than $\sigma_{\mathrm{A,rep}}$ for times $\geq1$s. The out-of-loop
OBP filter is centered approximately $n_{\mathrm{ol}}\sim73$ comb
lines away from the central pump line. We can write $f_{\mathrm{ol}}^{\mathrm{k}}=\Delta n_{\mathrm{ol}}\times f_{\mathrm{rep}}^{\mathrm{k}}+f_{\mathrm{o}}^{\mathrm{k}}$
with $\mathrm{k}$ being the measurement number and where $\Delta n_{ol}$
is the mode number relative to the central comb line. and thus $\sigma_{\mathrm{A,ol}}^{2}=\sigma_{\mathrm{A,o}}^{2}+\Delta n_{\mathrm{ol}}^{2}\,\sigma_{\mathrm{A,rep}}^{2}+\Delta n_{\mathrm{ol}}\left\langle (f_{\mathrm{rep}}^{\mathrm{k+1}}-f_{\mathrm{rep}}^{\mathrm{k}})(f_{\mathrm{o}}^{\mathrm{k+1}}-f_{\mathrm{o}}^{\mathrm{k}})\right\rangle $
and since $\sigma_{\mathrm{A,o}}\ll\sigma_{\mathrm{A,rep}}$ for $\tau\geq1$
s we expect $\sigma_{\mathrm{A,ol}}\approx\Delta n_{\mathrm{ol}}\:\sigma_{\mathrm{A,rep}}$
which in good agreement with the observed offset.

\begin{figure}[t!]
\centering
\includegraphics[width=0.85\columnwidth]{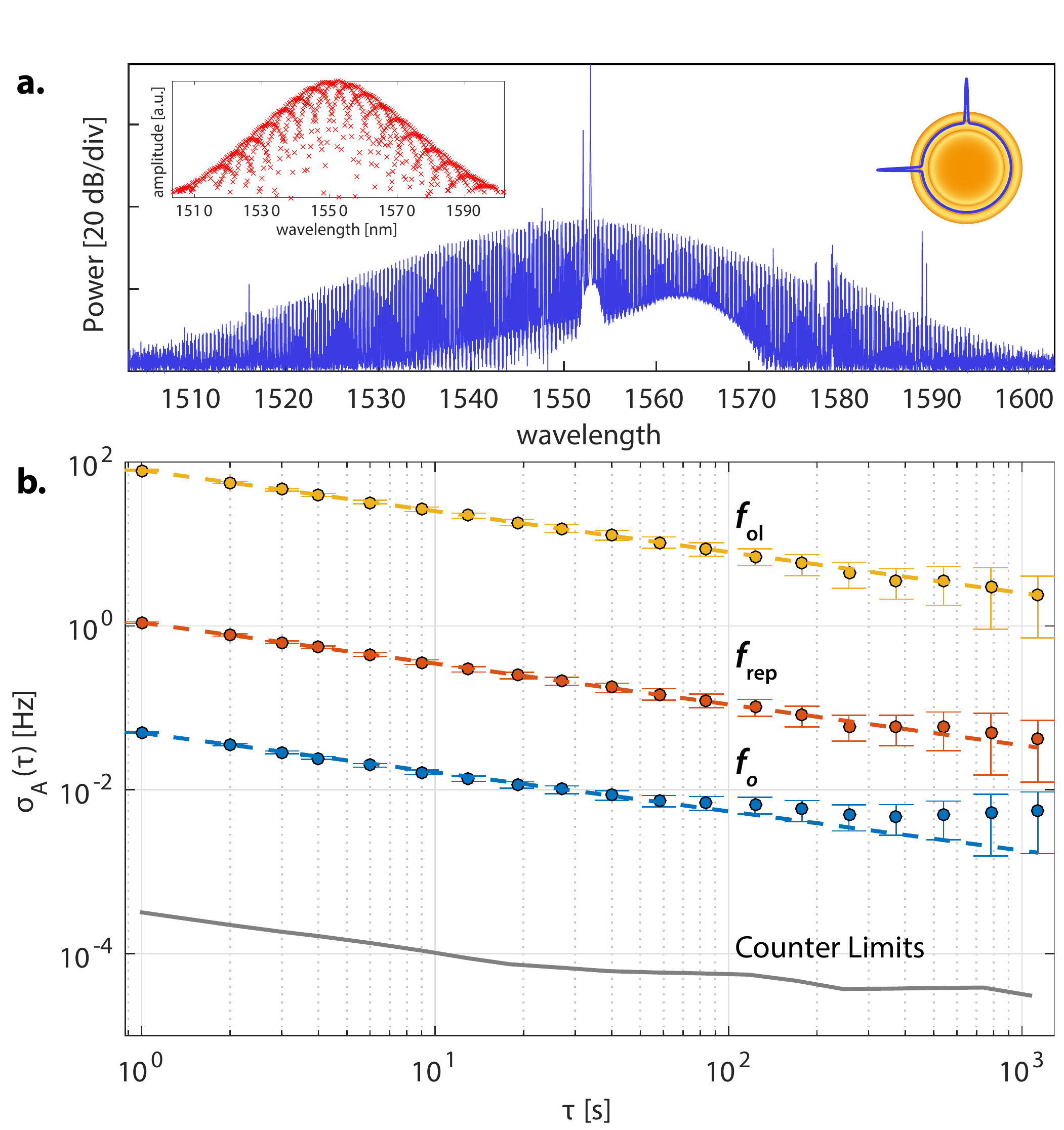}\protect\caption{Multi-soliton spectrum and overlapping Allan deviation. \textbf{a.}
A measurement of the optical spectrum from the OSA of the stabilized
two soliton state (blue). The peaks of the spectrum are detected and
fitted with a theoretical expression \cite{Brasch2014} (left inset).
This yields the duration, the number and relative position of the
solitons in the cavity (see the inset in the upper right). \textbf{b.}
Counting data at a 1 s gate time was taken for 3000 s. The resulting
OAD is shown by the traces for the pump frequency (blue), repetition
rate (red), and the independent out-of-loop measurement (yellow).
The error bars are one standard deviation. The dashed lines are a
linear fit to the data. The gray trace shows the result of counting
the 20 MHz master oscillator used in the experiment. \label{fig:Multi-Soliton-Spectrum}}
\end{figure}

We also demonstrate it is possible to fully stabilize multi-soliton
states. The desired state is created using the same tuning technique
\cite{Herr2013} for the single soliton. The spectrum for the created
state is shown in figure~\ref{fig:Multi-Soliton-Spectrum}. Having
multiple solitons alters the microresonator optical spectrum as a
result of interference between the different frequency components.
The time domain pattern of the intracavity amplitude can be reconstructed
by fitting an analytical expression to the optical spectrum \cite{Brasch2014}.
The left inset in figure~\ref{fig:Multi-Soliton-Spectrum} shows
only the peak amplitudes for clarity. It is estimated that the state
in figure~\ref{fig:Multi-Soliton-Spectrum} is a two soliton state
spaced apart by $\sim90^{\circ}$ in the microresonator. The same
stabilization was applied as for the single soliton with $\sim3000$~s
of counting data taken at 1 s gate times and the OAD was calculated
for $f_{\mathrm{rep}}$, $f_{\mathrm{o}}$ , and $f_{\mathrm{ol}}$,
shown in figure~\ref{fig:Multi-Soliton-Spectrum}b. At 1 s, the absolute
fluctuations of rep rate beat is $\sigma_{\mathrm{A,rep}}=1.1$~Hz
corresponding to a fractional deviation of $7.8\times10^{-11}$. The
absolute frequency fluctuations of $f_{\mathrm{o}}$ at 1 s are $\sigma_{\mathrm{A,o}}=5\times10^{-2}$~Hz.
For the $f_{\mathrm{ol}}$ the beat frequency was 48.7~MHz and the
absolute fluctuations at 1 s were found to be $\sigma_{\mathrm{A,ol}}=79.5$~Hz.
This is a factor of $\sim72.6$ higher than fluctuation seen in $f_{\mathrm{rep}}$,
which again agrees well with the out-of-loop filter position. The
bump in $\sigma_{\mathrm{A,rep}}$ and $\sigma_{\mathrm{A,ol}}$ is
absent here as the gate time was taken to be 1 s. Fitting the data
gives the slopes $\tau_{\mathrm{rep}}^{-0.50}$, $\tau_{\mathrm{o}}^{-0.48}$
and $\tau_{\mathrm{ol}}^{-0.50}$. This data has a lower initial OAD.
The servo parameters were changed between the single and multi-soliton
case, which potentially explains the difference.

In summary we have demonstrated that an auxiliary laser can be used
as actuator to stabilize the optical frequency comb spectra associated
with a temporal optical soliton in a microresonator. The distinct
advantage of this actuator is that it is easily implemented, and that
in addition the added noise is only limited to laser noise. Our technique
is applicable to a wide range of other microresonator platforms.

\subsection*{Funding}

This work was supported by a Marie Curie IIF (J. D. J.), the Swiss
National Science Foundation (T. H.), the European Space Agency (V.
B.), a Marie Curie IEF (C. L.), the Eurostars program, and the Defense
Advanced Research Program Agency (DARPA) PULSE program, grant number
W31P4Q-13-1-0016.

\subsection*{Acknowledgments}

$^{\dagger}$These authors contributed equally to this work. 

\bibliographystyle{ieeetr}
\bibliography{library}

\end{document}